%% file: main.tex
\documentclass[sigplan,10pt,dvipsnames,ocgcolorlinks,nonacm]{acmart}

\usepackage{balance}
\renewcommand\footnotetextcopyrightpermission[1]{} 

\setcopyright{none}

\setcitestyle{numbers,sort&compress}

\usepackage{rotating}

\usepackage{booktabs}   
\usepackage{subcaption} 

\usepackage{oz}
\usepackage{mathpartir}
\usepackage{FiraSans}
\usepackage[scaled=0.9]{FiraMono}
\usepackage[T1]{fontenc}

\usepackage{enumitem}

\usepackage{xspace}
\newcommand{\Lift}{\textsc{Lift}\xspace}
\newcommand{\Rise}{\textsf{\textsc{Rise}}\xspace}

\newcommand{\Elevate}{\textmd{\textsf{\textsc{\firalight Elevate}}}\xspace}
\newcommand{\FSmooth}{$\widetilde{\text{\textbf{F}}}$\xspace}

\definecolor{solarized-green}{RGB}{133,153,0}
\definecolor{solarized-blue}{RGB}{38,139,210}
\definecolor{solarized-magenta}{RGB}{211,54,130}

\usepackage{listings}

\usepackage{tikz}
\usetikzlibrary{snakes,arrows,shapes}
\usepackage{pgfplots}
\pgfplotsset{compat=1.11,
    /pgfplots/ybar legend/.style={
    /pgfplots/legend image code/.code={%
       \draw[##1,/tikz/.cd,yshift=-0.25em]
        (0cm,0cm) rectangle (3pt,0.8em);},
   },
}
\usepgfplotslibrary{colorbrewer}

\usepackage{cleveref}

\lstdefinestyle{halide}{
    language=c,
	keywordstyle=\bfseries\color{black!75},
  morekeywords= [2]{
    Func,RDom,
    bound,tile,split,vectorize,reorder,unroll,update,compute_at,store_in,
    gpu_blocks,gpu_threads,gpu_lanes,parallel,
    args,in
  },
  keywordstyle= [2]{\bfseries\color{MidnightBlue!75}},
  morekeywords = [3]{
    warp_size, vec_size,
    x_tile, y_tile,
    y_unroll, r_unroll
  },
  keywordstyle = [3]{\itshape\color{Violet}},
  morekeywords = [4]{
    x,y,r,z,xc,yc,
    xi,yi,xio,xii,yii,xo,yo,x_pair,xiio,ty,rxo,rxi,k,ki,ko,
    Bx,By,Ax,Ay
  },
  keywordstyle = [4]{\color{RedViolet}},
	basicstyle=\normalsize\ttfamily\scriptsize, 
	commentstyle=\itshape\color{gray}, 
  stringstyle=\itshape, 
  xleftmargin=2.5em,
	numbers=left, 
	numberstyle=\scriptsize, 
	stepnumber=1, 
    tabsize = 2,
	numbersep=8pt, 
	showstringspaces=false, 
	breaklines=true, 
	backgroundcolor = \color{black!05},
  frame=lines, 
	abovecaptionskip=.5\baselineskip, 
	aboveskip=\baselineskip,
  escapechar=@,
	captionpos=b,
	mathescape=true,
}

\lstdefinestyle{halide-short}{
    language=c,
	keywordstyle=\bfseries\color{black!75},
  morekeywords= [2]{
    Func,RDom,
    bound,tile,split,vectorize,reorder,unroll,update,compute_at,store_in,
    gpu_blocks,gpu_threads,gpu_lanes,parallel,
    args,in
  },
  keywordstyle= [2]{\bfseries\color{MidnightBlue!75}},
  morekeywords = [3]{
    warp_size, vec_size,
    x_tile, y_tile,
    y_unroll, r_unroll
  },
  keywordstyle = [3]{\itshape\color{Violet}},
  morekeywords = [4]{
    x,y,r,z,xc,yc,
    xi,yi,xio,xii,yii,xo,yo,x_pair,xiio,ty,rxo,rxi,k,ki,ko,
    Bx,By,Ax,Ay
  },
  keywordstyle = [4]{\color{RedViolet}},
	basicstyle=\normalsize\ttfamily\scriptsize, 
	commentstyle=\itshape\color{gray}, 
  stringstyle=\itshape, 
  numbers=none, 
  tabsize = 2,
  xleftmargin=.5\parindent,
	numbersep=8pt, 
	showstringspaces=false, 
	breaklines=true, 
  aboveskip=.5em,
  belowskip=.5em,
  escapechar=@,
	captionpos=b,
	mathescape=true,
}
\newcommand{\halideInline}[1]{\lstinline[style=halide-short]!#1!}

\usepackage{upquote}

\lstdefinelanguage{FSharp}%
{morekeywords={let, new, match, with, rec, open, module, namespace, type, of, member, %
and, for, while, true, false, in, do, begin, end, fun, function, return, yield, try, %
mutable, if, then, else, cloud, async, static, use, abstract, interface, inherit, finally },
  otherkeywords={ let!, return!, do!, yield!, use!, var, from, select, where, order, by },
  keywordstyle=\color{bluekeywords},
  sensitive=true,
  basicstyle=\ttfamily,
	breaklines=true,
  xleftmargin=\parindent,
  aboveskip=\bigskipamount,
	tabsize=4,
  morecomment=[l][\color{greencomments}]{///},
  morecomment=[l][\color{greencomments}]{//},
  morecomment=[s][\color{greencomments}]{{(*}{*)}},
  morestring=[b]",
  showstringspaces=false,
  literate={`}{\`}1,
  stringstyle=\color{redstrings},
}

\lstdefinestyle{fsmooth}{
  language=FSharp,
	keywordstyle=\bfseries\color{black!75},
	basicstyle=\normalsize\ttfamily\footnotesize, 
	commentstyle=\itshape\color{gray}, 
  stringstyle=\itshape, 
  xleftmargin=2.5em,
	numbers=left, 
	numberstyle=\scriptsize, 
	stepnumber=1, 
    tabsize = 2,
	numbersep=8pt, 
	showstringspaces=false, 
	breaklines=true, 
	backgroundcolor = \color{black!05},
  frame=lines, 
	abovecaptionskip=.5\baselineskip, 
	aboveskip=\baselineskip,
  escapechar=@,
	captionpos=b,
	mathescape=true,
}

\lstdefinestyle{fsmooth-short}{
  language=FSharp,
	keywordstyle=\bfseries\color{black!75},
	basicstyle=\normalsize\ttfamily\footnotesize, 
	commentstyle=\itshape\color{gray}, 
  stringstyle=\itshape, 
  xleftmargin=.5\parindent,
	numbersep=8pt, 
	showstringspaces=false, 
	breaklines=true, 
	abovecaptionskip=.5\baselineskip, 
	aboveskip=.5em,
  belowskip=.5em,
  escapechar=@,
	captionpos=b,
	mathescape=true,
}

\lstdefinelanguage{Lift}
{
morekeywords={
        map, split, slide, transpose, def, T, *, S
},
}

\lstdefinestyle{tvm}{
  language=Python,
  morekeywords = {tvm, lambda,compute,sum,indexmod},
  morekeywords= [2]{
    tile,
    split,
    reorder,
    vectorize,
    parallel,
    unroll,
    compute_at
  },
  keywordstyle= [2]{\bfseries\color{MidnightBlue!75}},
  morekeywords = [3]{
    name,
    axis
      },
  keywordstyle = [3]{\itshape\color{Violet}},
  morekeywords = [4]{
    0, K, k, 'k', M, N, x, y, A, B, C, s, 32, xo, yo, xi, yi, i, j, ko, ki,
    m, n, mTile, nTile, row, col, z, packedB, CC, xc, yc
  },
  keywordstyle = [4]{\color{RedViolet}},
  keywordstyle=\bfseries\color{black!75}, 
   	basicstyle=\ttfamily\scriptsize, 
	commentstyle=\color{gray}, 
	stringstyle=\itshape, 
	xleftmargin=2.5em,
	numbers=left, 
	numberstyle=\scriptsize, 
	stepnumber=1, 
  tabsize = 2,
	numbersep=8pt, 
	showstringspaces=false, 
	breaklines=true, 
	backgroundcolor = \color{black!05},
	frame=lines, 
	abovecaptionskip=.5\baselineskip, 
	aboveskip=\baselineskip,
  escapechar=@,
	captionpos=b,
  mathescape=true,
  literate={`}{\lq}1 
}

\lstdefinestyle{elevate-rise}{
  language=scala,
  morekeywords = {apply},
  deletekeywords={try},
  otherkeywords = {;, |>, =>, >>, <+},
  keywordstyle=\bfseries\color{black!75}, 
  morekeywords= [2]{ 
  ;,seq,
  <+,lChoice,
  try,
  repeat,
  all,one,some,
  topDown,
  allTopDown,
  tryAll,
  allBottomUp,
  bottomUp,
  normalize,
  Strategy,
  RewriteResult,
  Success,
  Failure
  },
  keywordstyle= [2]{\bfseries\color{MidnightBlue!75}},
  morekeywords = [3]{ 
    separateDot,
    lowerToSeqC,
    mapFusion,
    fuseReduceMap,
    isReduce,
    isMap,
    lowerToC,
    blocking,
    loopPerm,
    arrayPacking,
    packB,
    parallelizeCopy,
    body,
    function,
    argument,
    fmap,
    reorder,
    interchange,
    argOf,
    not,isFun,
    betaReduction,
    etaAbstraction,etaReduction,
    id,fail,
    buildGet, lenBuild, letPartialEval, letApp, funToLet,
    parallel,vectorize,unroll,tile,tileND,DFNF,BENF
  },
  keywordstyle = [3]{\color{RoyalPurple}},
  morekeywords = [4]{ 
    p, e, s
  },
  keywordstyle = [4]{\color{Plum}},
  morekeywords = [5]{
    bf,
    mm,
    dot,
    threemaps,
    mt
  },
  keywordstyle = [5]{\bfseries\color{RedViolet}},
  morekeywords= [6]{ 
    |>, >>,
    fun,
    app,
    map,
    zip,
    reduce,reduceSeq,reduceSeqUnroll,
    mapSeq,mapSeqUnroll,
    mapSeq,
    mapVec,
    asVector,
    asScalar,
    map2D,
    pad,
    pad2D,
    slide,
    slide2D,
    join,
    transpose,
    split
  },
  keywordstyle= [6]{\bfseries\color{OliveGreen!95}},
  morekeywords = [7]{ 
    dot,
    mm
  },
  keywordstyle = [7]{\color{RawSienna}},
  morekeywords = [8]{ 
    weights2d,weightsH,weightsV,
    nbh,
    x,y,
    f,nf,
    g,
    xs,
    img,
    a, na,
    b, nb,
    M,K,N,
    ak, bk,arow,bcol,
    float,
  },
  keywordstyle = [8]{\color{RedOrange}},
	basicstyle=\ttfamily\scriptsize, 
	commentstyle=\itshape, 
	stringstyle=\itshape, 
	xleftmargin=2.5em,
	numbers=left, 
	numberstyle=\scriptsize, 
	stepnumber=1, 
  tabsize = 2,
	numbersep=8pt, 
	showstringspaces=false, 
	breaklines=true, 
	backgroundcolor = \color{black!05},
	frame=lines, 
	abovecaptionskip=.5\baselineskip, 
	aboveskip=\baselineskip,
  escapechar=@,
	captionpos=b,
  mathescape=true,
  literate={`}{\lq}1 
}

\lstdefinestyle{elevate-rise-short}{
  language=scala,
  morekeywords = {apply, List},
  deletekeywords={try},
  otherkeywords = {;, |>, =>, >>, <+},
  keywordstyle=\bfseries\color{black!75}, 
  morekeywords= [2]{ 
    ;,seq,
    <+,lChoice,
    try,
    repeat,
    all,one,some,
    topDown,
    allTopDown,
    tryAll,
    allBottomUp,
    bottomUp,
    normalize,
    Strategy,
    RewriteResult,
    Success,
    Failure
  },
  keywordstyle= [2]{\bfseries\color{MidnightBlue}},
  morekeywords = [3]{ 
    separateDot,
    lowerToSeqC,
    mapFusion,
    fuseReduceMap,
    isReduce,
    isMap,
    lowerToC,
    blocking,
    loopPerm,
    body,
    function,
    argument,
    fmap,
    interchange,
    reorder,
    argOf,
    not,isFun,
    betaReduction,
    etaAbstraction,etaReduction,
    id,fail,
    buildGet, lenBuild, letPartialEval, letApp, funToLet,
    parallel,vectorize,unroll,tile,tileND,DFNF,BENF
  },
  keywordstyle = [3]{\color{RoyalPurple}},
  morekeywords = [4]{ 
    p, e, s, fs, ss
  },
  keywordstyle = [4]{\color{Plum}},
  morekeywords = [5]{
    bf,
    mm,
    threemaps,
    mt
  },
  keywordstyle = [5]{\bfseries\color{RedViolet}},
  morekeywords= [6]{ 
  |>, >>,
  fun,
  app,
  map,
  zip,
  reduce,reduceSeq,reduceSeqUnroll,
  mapSeq,mapSeqUnroll,
  mapPar,
  mapVec,
  asVector,
  asScalar,
  map2D,
  pad,
  pad2D,
  slide,
  slide2D,
  join,
  transpose,
  split
  },
  keywordstyle= [6]{\bfseries\color{OliveGreen}},
  morekeywords = [7]{ 
    dot
  },
  keywordstyle = [7]{\color{RawSienna}},
  morekeywords = [8]{ 
    weights2d,weightsH,weightsV,
    nbh,
    x,y,
    f,nf,
    g,
    xs,
    img,
    a, na,arow,bcol,
    b, nb
  },
  keywordstyle = [8]{\color{RedOrange}},
	basicstyle=\ttfamily\footnotesize, 
	commentstyle=\itshape, 
	stringstyle=\itshape, 
	numbers=none, 
  tabsize = 2,
  xleftmargin=.5\parindent,
	numbersep=8pt, 
	showstringspaces=false, 
	breaklines=true, 
  aboveskip=.5em,
  belowskip=.5em,
  escapechar=@,
	captionpos=b,
  mathescape=true,
  literate={`}{\lq}1 
}
\newcommand{\inline}[1]{\lstinline[style=elevate-rise-short]!#1!}
\newcommand{\elevateUserFun}[1]{\lstinline[basicstyle=\ttfamily\footnotesize\color{RoyalPurple}]!#1!}

\definecolor{light-gray}{gray}{0.85}

\begin{document}

\title[\Elevate: A Language for Describing Optimization Strategies]{A Language for Describing Optimization Strategies}         


\author{Bastian Hagedorn}
\affiliation{
  \institution{University of M{\"u}nster, Germany}            
}
\email{b.hagedorn@wwu.de}          

\author{Johannes Lenfers}
\affiliation{
  \institution{University of M{\"u}nster, Germany}            
}
\email{j.le@wwu.de}          

\author{Thomas Koehler}
\affiliation{
  \institution{University of Glasgow, UK}           
}
\email{t.koehler.1@research.gla.ac.uk}          
\author{Sergei Gorlatch}
\affiliation{
  \institution{University of M{\"u}nster, Germany}            
}
\email{gorlatch@wwu.de}          
\author{Michel Steuwer}
\orcid{0000-0001-5048-0741}             
\affiliation{
  \institution{University of Glasgow, UK}           
}
\email{michel.steuwer@glasgow.ac.uk}         

\begin{abstract}

    Optimizing programs to run efficiently on modern parallel hardware is hard but crucial for many applications.
    The predominantly used imperative languages - like C or OpenCL - force the programmer to intertwine the code describing functionality and optimizations.
    This results in a nightmare for portability which is particularly problematic given the accelerating trend towards specialized hardware devices to further increase efficiency.


    Many emerging DSLs used in performance demanding domains such as deep learning, automatic differentiation, or image processing attempt to simplify or even fully automate the optimization process.
    Using a high-level - often functional - language, programmers focus on describing functionality in a declarative way.
    In some systems such as Halide or TVM, a separate \emph{schedule} specifies how the program should be optimized.
    Unfortunately, these schedules are not written in well-defined programming languages.
    Instead, they are implemented as a set of ad-hoc predefined APIs that the compiler writers have exposed.


  In this paper, we present \emph{\Elevate}: a functional language for describing optimization strategies.
  \Elevate follows a tradition of prior systems used in different contexts that express optimization strategies as composition of rewrites.
  In contrast to systems with scheduling APIs, in \Elevate programmers are not restricted to a set of built-in optimizations but define their own optimization strategies freely in a composable way.
  We show how user-defined optimization strategies in \Elevate enable the effective optimization of programs expressed in a functional data-parallel language demonstrating competitive performance with Halide and TVM.
\end{abstract}




\maketitle

\section{Introduction}
\input{tex/intro}

\section{Motivation and Background}
\label{sec:motivation}
\input{tex/motivation}

\section{\Elevate{}: A Language for Describing Optimization Strategies}
\label{sec:elevate}
\input{tex/elevate}

\section{Case Study 1: Automatic Differentiation}
\label{sec:case-study-1}
\input{tex/fsmooth}

\section{Case Study 2: Image Processing}
\label{sec:case-study-2}
\input{tex/halide}

\section{Case Study 3: Deep Learning}
\label{sec:case-study-3}
\input{tex/tvm}



\section{Related Work}
\label{sec:related-work}
\input{tex/related}

\section{Conclusion}
\label{sec:conclusion}
\input{tex/conclusion}

\begin{acks}                            

We would like to thank the entire \Rise (rise-lang.org) and \Elevate (elevate-lang.org) teams for their development efforts.
The first author was financially supported by an Nvidia Research Fellowship.
\end{acks}

\balance
\bibliography{bibfile}



\end{document}

%% file: tex/intro.tex
The tremendous gains in performance and efficiency that computer hardware continues to make are a key driving force for innovation in computing.
This enables entire new areas of computing such as deep learning to deliver applications unthinkable even just a few years ago.
With the end of Moore's law and Denard's scaling~\citep{DBLP:journals/cacm/HennessyP19}, these gains no longer come for free for software writers.
Programs have to be optimized for an increasing diverse set of hardware devices by exploiting many subtle details of the computer architecture.
Performance portability has emerged as a crucial concern as software naturally outlives the faster cycle of hardware generations.
In addition, specialized hardware has proven to offer extreme benefits for performance and energy efficiency - if the specially optimized software exploits it.

The predominant imperative and low-level programming approaches such as C, CUDA, or OpenCL force programmers to intertwine the code describing the functional behavior of the program with optimization decisions.
This makes them -- by design -- non performance portable.
As an alternative, higher level domain-specific approaches have emerged that allow programmers to declaratively describe the functional behavior without committing to a specific implementation.
Popular examples of this approach are virtually all machine learning systems such as TensorFlow~\citep{tensorflow2015-whitepaper} or PyTorch~\citep{paszke2017automatic}.
For these approaches, the compilers and runtime systems are responsible to optimize the computations that are expressed as data-flow graphs.
Programmers have limited control about the optimization process.
Instead large teams of engineers at Google and Facebook provide fast implementations for the most common hardware platforms, for TensorFlow including Google's specialized TPU hardware.
This labour intensive support of new hardware devices is currently only sustainable for the biggest companies in the market -- and even they struggle~\cite{DBLP:conf/hotos/0001I19}.
To overcome this innovation obstacle and to achieve automated performance portability we will need to rethink how we separate, describe, and apply optimizations in a more principled way.

Encoding program transformations as \emph{rewrite rules} has been a long established idea.
\citet{Bird:1997:AP:248932} studied an algebraic programming approach where functional programs are rewritten by exploiting algebraic properties.
The Glasgow Haskell Compiler allows the specification of rewrite rules for program optimizations~\cite{peytonjones2001playing}.
More recently, \Lift~\cite{DBLP:conf/icfp/SteuwerFLD15} encodes optimization and implementation choices as rewrite rules for optimizing a high-level pattern-based data-parallel functional language using an automated stochastic search method applying the rewrites.
Rewrite based approaches, such as \Lift, have the advantage of being easily extensible towards new application domains (such as stencils~\cite{DBLP:conf/cgo/HagedornSSGD18}) as well as supporting new hardware features (such as specialized vector instructions which are encoded as new low-level patterns and introduced by a rewrite rule~\cite{DBLP:conf/cases/SteuwerRD16}).
Unfortunately, these rewrite approaches are limited in their practicality to deliver the high performance required in many real-world applications.
They lack control over the rewriting process and the automated rewriting using stochastic search processes takes a long time to find a high performance implementation.
In this paper, we are going to address these practical limitations of rewrite-based approaches for optimizing high-performance real-world applications by defining a strategy language that allows the definition of optimization strategies that precisely controls the rewrite process.

Halide~\cite{DBLP:conf/pldi/Ragan-KelleyBAPDA13, cacm/Ragan-KelleyASB18} has introduced the concept of separating programs into functional descriptions and schedules in the area of high-performance domain-specific code generators.
A \emph{schedule} describes the optimizations to be applied to the Halide \emph{algorithm} that defines the functional behavior of the computation.
Halide's schedules -- as well similar schedules in TVM~\cite{DBLP:conf/osdi/ChenMJZYSCWHCGK18} -- are implemented using a set of predefined APIs that expose a fixed set of optimization options.
Halide's authors describe these APIs as a scheduling \emph{language} but it lacks many desirable properties of a programming language.
Most crucially, programmers are not able to define their own abstractions.
Even the composition of existing optimization primitives is in some cases unintuitive due to the lack of a clear semantics and Halide's compiler has default and implicit behavior limiting experts' control.
All of these reasons make writing schedules in Halide significantly harder than writing algorithms.
Furthermore, for some desirable optimizations it is not sufficient to change the schedule but the algorithm itself has to be redefined -- violating the promise of separating algorithm and schedule.
In this paper, we build upon Halide's general idea but provide a proper functional \emph{strategy language}, called \Elevate, with clear semantics of individual primitives and how they compose.
It enables programmers to define their own abstractions for building optimization strategies in a composable and reusable way.

The design of \textit{\Elevate{}} is heavily inspired by research on strategy languages for rewrite systems used in other contexts -- and largely unknown to the high-performance code generation community -- such as Stratego~\cite{rta/Visser01}.
\citet{birthday/Kirchner15} provides a recent overview of the research of the rewriting community.
We claim no novelty in the design foundations of strategy languages but instead in the strategies we present and their usage to facilitate the generation of highly efficient code on modern hardware.

Our paper makes the following key contributions:
\begin{itemize}
    \item Description of the design of \Elevate{}, a functional language for describing optimization strategies for high-performance code generation (\Cref{sec:elevate});
    \item demonstration of \Elevate{} using three case studies: automatic differentiation (\cref{sec:case-study-1}), image processing (\cref{sec:case-study-2}) and deep learning (\cref{sec:case-study-3}).
          They show the \emph{flexibility} and \emph{extensibility} of \Elevate{} and experimentally evaluate the \emph{practicality} of a rewrite based approach for achieving competitive high performance.
\end{itemize}



%% file: tex/motivation.tex

We motivate the need for a strategy language with a closer look at Halide.
We then argue for a more principled language approach for describing optimizations strategies. 


\subsection{Halide: Decoupling Algorithm from Schedules}
Halide~\cite{cacm/Ragan-KelleyASB18} has originally been designed to generate high performance code for image processing pipelines~\citep{DBLP:conf/pldi/Ragan-KelleyBAPDA13}, but has since inspired similar approaches in other contexts such as TVM in deep leaning~\citep{DBLP:conf/osdi/ChenMJZYSCWHCGK18}.
A crucial idea is the separation of a program in two parts:
the \emph{algorithm} describing the functional behavior, and
the \emph{schedule} specifying how the program should be optimized by the underlying Halide compiler.

\begin{lstlisting}[
  style=halide,
  float={b},
  caption={%
    Matrix matrix multiplcation in Halide. Lines~\ref{lst:halide-mm:computation:begin}--\ref{lst:halide-mm:computation:end} define the computation $A \times B$, the other lines define the schedule specifying the optimizations to be applied by the compiler.
    {\scriptsize From: { \url{https://github.com/halide/Halide/blob/master/apps/cuda_mat_mul/mat_mul_generator.cpp}}}.
  },
  label={lst:halide-mm}
]
// functional description of matrix multiplication
Var x("x"), y("y"); Func prod("prod"); RDom r(0, size);$\label{lst:halide-mm:computation:begin}$
prod(x, y) += A(x, r) * B(r, y);$\label{lst:halide-mm:computation}$
 out(x, y)  = prod(x, y);$\label{lst:halide-mm:computation:end}$

// schedule for Nvidida GPUs
const int warp_size = 32; const int vec_size = 2;
const int x_tile    =  3; const int y_tile   = 4;
const int y_unroll  =  8; const int r_unroll = 1;
Var xi,yi,xio,xii,yii,xo,yo,x_pair,xiio,ty; RVar rxo,rxi;
out.bound(x, 0, size).bound(y, 0, size)
    .tile(x, y, xi, yi, x_tile * vec_size * warp_size,
          y_tile * y_unroll)
    .split(yi, ty, yi, y_unroll)
    .vectorize(xi, vec_size)
    .split(xi, xio, xii, warp_size)
    .reorder(xio, yi, xii, ty, x, y)
    .unroll(xio).unroll(yi)
    .gpu_blocks(x, y).gpu_threads(ty).gpu_lanes(xii);
prod.store_in(MemoryType::Register).compute_at(out, x)
    .split(x, xo, xi, warp_size * vec_size, RoundUp)$\label{lst:halide-mm:repeat1:start}$
    .split(y, ty, y, y_unroll)
    .gpu_threads(ty).unroll(xi, vec_size).gpu_lanes(xi)$\label{lst:halide-mm:repeat1:end}$
    .unroll(xo).unroll(y).update()
    .split(x, xo, xi, warp_size * vec_size, RoundUp)$\label{lst:halide-mm:repeat2:start}$
    .split(y, ty, y, y_unroll)
    .gpu_threads(ty).unroll(xi, vec_size).gpu_lanes(xi)$\label{lst:halide-mm:repeat2:end}$
    .split(r.x, rxo, rxi, warp_size)
    .unroll(rxi, r_unroll).reorder(xi, xo, y, rxi, ty, rxo)
    .unroll(xo).unroll(y);
Var Bx = B.in().args()[0], By = B.in().args()[1];
Var Ax = A.in().args()[0], Ay = A.in().args()[1];
B.in().compute_at(prod, ty).split(Bx, xo, xi, warp_size)
      .gpu_lanes(xi).unroll(xo).unroll(By);
A.in().compute_at(prod, rxo).vectorize(Ax, vec_size)
      .split(Ax,xo,xi,warp_size).gpu_lanes(xi).unroll(xo)
      .split(Ay,yo,yi,y_tile).gpu_threads(yi).unroll(yo);
A.in().in().compute_at(prod, rxi).vectorize(Ax, vec_size)
      .split(Ax, xo, xi, warp_size).gpu_lanes(xi)
      .unroll(xo).unroll(Ay);
\end{lstlisting}

\Cref{lst:halide-mm} shows a snippet of Halide code used for generating an efficient matrix-matrix multiplication for an Nvidia GPU.
Halide is a DSL embedded in C++, so the syntax used here is C++.
The lines~\ref{lst:halide-mm:computation:begin}--\ref{lst:halide-mm:computation:end} define the matrix-matrix multiplication computation:
$A$ and $B$ are multiplied by performing the dot product for each coordinate pair $(x,y)$.
The dot product is expressed as pairwise multiplications and reducing over the reduction domain $r$ using the \texttt{+=} operator (line~\ref{lst:halide-mm:computation}).

The other lines in the listing define the schedule specifying the optimizations to be performed.
The Halide compiler takes this C++ program and produces efficient GPU code coming close to highly optimized low-level library code.

By looking at the code it is immediately clear that writing a schedule is significantly more challenging than writing the algorithm describing matrix-matrix multiplication.
Schedules are written using a sequence of API calls on the C++ objects that represent the input (\texttt{A}, \texttt{B}) and output (\texttt{out}) data.
\texttt{prod} represents the reduction operation in Halide's internal representation.
While the algorithm and schedule are separated they still share the same C++ identifiers and must, therefore, be written in the same C++ scope limiting the reuse of schedules across algorithms.

This schedule uses 12 built-in optimization primitives ({\small\texttt{bound}, \texttt{tile}, \texttt{split}, \texttt{vectorize}, \texttt{reorder}, \texttt{unroll}, \texttt{update}, \texttt{compute\_at}, \texttt{store\_in}, \texttt{gpu\_blocks}, \texttt{gpu\_threads}, \texttt{gpu\_lanes}}).
Some of these optimizations are specific for the hardware (like {\small\texttt{vectorize}} or {\small\texttt{gpu\_threads}}), others are generally useful algorithmic optimizations for many applications (like {\small\texttt{tiling}} to increase data locality), and others are low-level optimizations (like {\small\texttt{unroll}} and {\small\texttt{reorder}} that transform loop nests).
Halide is not easily extensible.
Adding a new optimization primitive to the schedule API requires extending the Halide compiler.
Even a primitive like {\small \texttt{tile}} that can be implemented with {\small \texttt{split}} and {\small\texttt{reorder}}\footnote{See: {\scriptsize\url{https://halide-lang.org/tutorials/tutorial_lesson_05_scheduling_1.html}}} is represented as a composition but provided as a built-in abstraction.
Halide's schedules lack the ability for user-defined abstractions.

The behavior of some primitives is not intuitive and the documentation provides only informal descriptions, e.g., for {\small\texttt{update}}: \textit{``Get a handle on an update step for the purposes of scheduling it''}.
The lack of clear descriptions of the optimization primitives makes reasoning about the schedule difficult.
For example, is it unclear to us why lines \ref{lst:halide-mm:repeat1:start}--\ref{lst:halide-mm:repeat1:end} are repeated at lines \ref{lst:halide-mm:repeat2:start}--\ref{lst:halide-mm:repeat2:end} with calls to {\small\texttt{unroll}} and {\small\texttt{update}} in between.


\medskip

\begin{figure}
	\begin{subfigure}{\linewidth}
\begin{lstlisting}[style=halide]
Var x, y; Func out; Func in = BC::repeat_edge(input);
out(x, y) = (
  1.f * in(x-1,y-1) + 2.f * in(x,y-1) + 1.f * in(x+1,y-1) +
  2.f * in(x-1,y)   + 4.f * in(x,y)   + 2.f * in(x+1,y)   +
  1.f * in(x-1,y+1) + 2.f * in(x,y+1) + 1.f * in(x+1,y+1)
) * (1.f/16.f);
\end{lstlisting}
	\end{subfigure}\vspace{-.75em}
	\begin{subfigure}{\linewidth}
\begin{lstlisting}[style=halide]
Var x,y; Func b_x,b_y,out; Func in=BC::repeat_edge(input);
b_y(x, y) =  in(x, y-1) + 2.f * in(x, y)  +  in(x, y+1);
b_x(x, y) = b_y(x-1, y) + 2.f * b_y(x, y) + b_y(x+1, y);
out(x, y) = b_x(x, y) * (1.f/16.f);
\end{lstlisting}
	\end{subfigure}
	\caption{Two-dimensional binomial filter in Halide (top) and separated version (bottom).}
	\label{lst:halide-bf}
\end{figure}

\Cref{lst:halide-bf} shows the implementation of a two-dimensional binomial filter in Halide.
The listing on top shows the basic Halide algorithm implementation by computing every output pixel as a weighted sum of the $3\times{}3$ surrounding pixels.
A more efficient way is shown below where the computation is separated into a vertical and horizontal filter operating on $3$ pixels each.
\emph{Separability} is a well known domain-specific optimization in image processing and only valid for appropriate weights.
Halide does not offer this optimization as a scheduling primitive and, therefore, programmers are forced to change the algorithm for this optimization -- clearly violating the promise of separating algorithm and schedule.

If no schedules are provided (as in \cref{lst:halide-bf}), the Halide compiler employs a set of implicit default optimizations that are out of reach of the control of the user.
This sometimes leads to the surprising behavior that algorithms without a schedule perform better (e.g., due to auto-vectorization) than ones where a schedule is provided.

\subsection{Towards an Optimization Strategy Language}
Out of the shortcoming of the API approach we identify the following desirable features for a strategy language:
\vspace{-.25em}
\begin{enumerate}[leftmargin=4mm]
	\item Optimization strategies should be defined clearly separated from the computational program facilitating reusablility of strategies across programs;
	\item Strategies should be written as compositions of user-defined strategies (possibly domain-specific ones); the language should facilitate the creation of higher-level abstractions;
	\item All strategies should have a clear semantics allowing reasoning about their application and implicit default behavior should be avoided to empower users to be in control.
\end{enumerate}
\vspace{-.25em}
\noindent
\emph{Essentially we argue that a strategy language should be build with the same standards as a language describing computation.}
In this paper we present such a language: \Elevate{}.

\begin{figure}
	\begin{subfigure}{\linewidth}
\begin{lstlisting}[style=elevate-rise]
val bf = fun(3.3.float)(weights2d => fun(N.M.float)(img =>
  img |> pad2D(1) |> slide2D(3)(1) |> map2D(fun(nbh =>
    dot(join(weights2d))(join(nbh)) ))))
\end{lstlisting}
	\end{subfigure}\vspace{-.75em}
	\begin{subfigure}{\linewidth}
\begin{lstlisting}[style=elevate-rise]
val sbf = (topDown(separateDot) `;` lowerToC)(bf)
\end{lstlisting}
	\end{subfigure}
	\caption{2D binomial filter in a \Lift-like language (top) and the separating optimization strategy in \Elevate (bottom).}
	\label{lst:elevate-bf}
\end{figure}


\Cref{lst:elevate-bf} shows an example of an \Elevate{} strategy (bottom) that optimizes a program (top) written in a completely separate pattern-based language similar to \Lift~\cite{DBLP:conf/cgo/HagedornSSGD18} called \Rise.
The optimization strategy is a sequential composition (\inline{`;`}) of user defined strategies providing a higher level of abstraction.
Strategies are build as compositions of rewrite rules with a clear semantics and no implicit behavior.

In the remainder of the paper, we will explain the language for writing such strategies and defining custom abstractions.

%% file: tex/elevate.tex
In this section, we describe our language for describing optimization strategies: \Elevate{}.
It is heavily inspired by earlier works on strategy languages for term rewriting systems, particularly Stratego~\cite{icfp/VisserBT98}.
Our key contribution is not the design of \Elevate itself but rather its application to define optimization for high-performance code generators.
In this section, we focus on introducing \Elevate as a practical programming language.
For a more formal treatment of strategy languages see~\cite{visser2004program}.

\subsection{Language Features and Types}
\Elevate is a functional language with a standard feature set including function recursion, algebraic data types and pattern matching.
Besides the standard scalar data types such as \inline{int}, types of interests are function types, tuple types and list types.
Our current implementation is an embedded DSL in Scala and we use Scala-like notation in the paper.

\subsection{Strategies}
A \emph{strategy} is the fundamental building block of \Elevate.
Strategies encode program transformations and are modeled as functions with the following type:
\begin{lstlisting}[style=elevate-rise-short]
type Strategy[P] = P => RewriteResult[P]
\end{lstlisting}
Here \inline{P} is the type of the program that is rewritten.
\inline{P} could for example be \inline{Rise} for programs written in the \Rise language from \cref{lst:elevate-bf}.
A \inline{RewriteResult[P]} is an algebraic datatype encoding the success or failure of applying a strategy to a program:
\begin{lstlisting}[style=elevate-rise-short]
RewriteResult[P] = Success[P](p: P)
                 | Failure[P](s: Strategy[P])
\end{lstlisting}
In case of a successful application, \inline{Success} contains the transformed program, in case of a failure, \inline{Failure} contains the strategy that has been unsuccessful.

The simplest example of a strategy is the \inline{id} strategy which always succeeds and returns its input program:
\begin{lstlisting}[style=elevate-rise-short]
def id[P]: Strategy[P] = (p: P) => Success(p)
\end{lstlisting}
The \inline{fail} strategy does the opposite and always fails while recording that the \inline{fail} strategy was the one failing:
\begin{lstlisting}[style=elevate-rise-short]
def fail[P]: Strategy[P] = (p: P) => Failure(fail)
\end{lstlisting}

\subsection{Rewrite Rules as Strategies}

\begin{figure}
	\includegraphics[width=.9\linewidth]{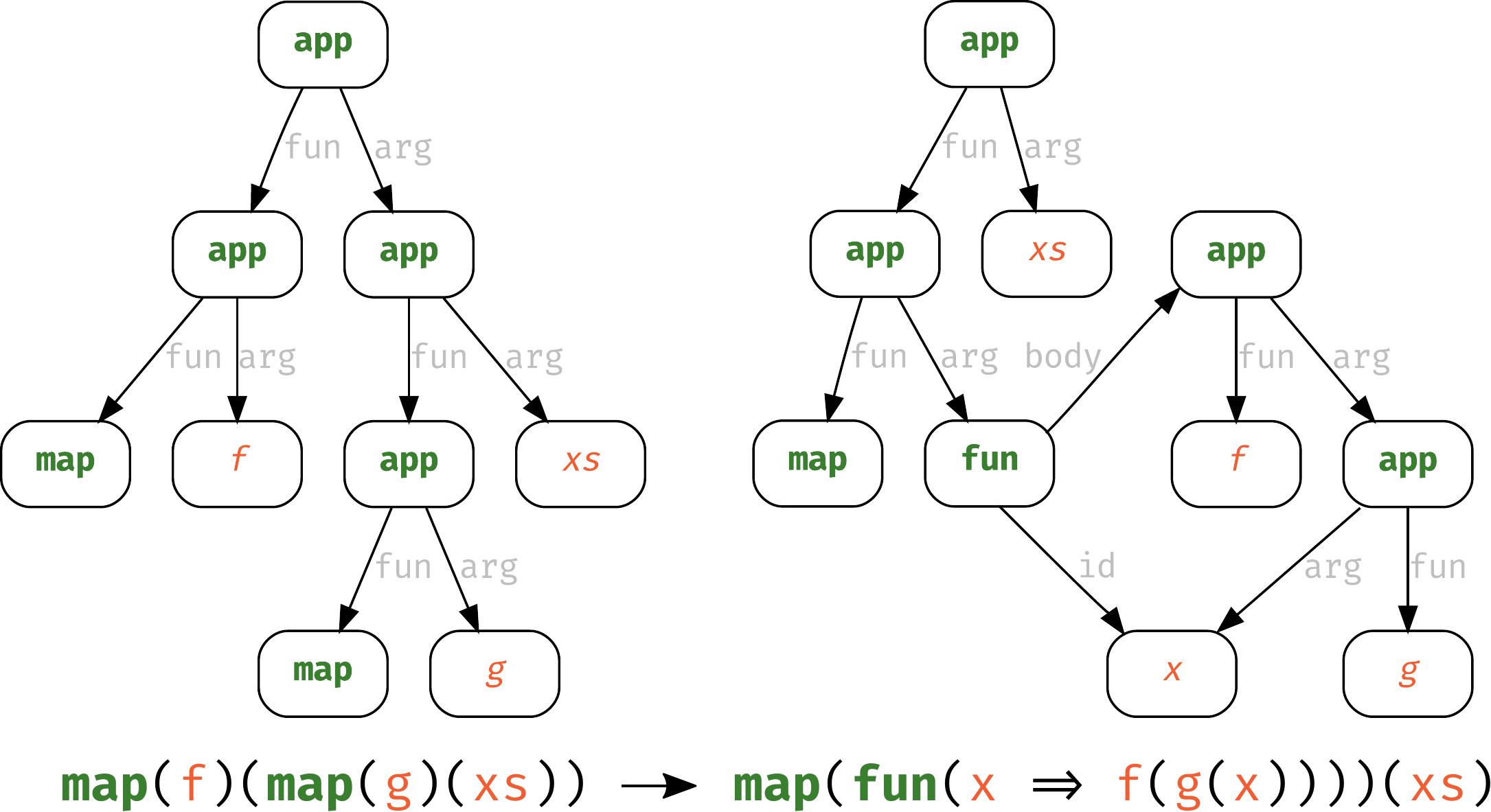}
        \caption{\Rise's \emph{map-fusion} rule as an AST transformation.} 
        \label{fig:mapFusion}
\end{figure}

In \Elevate, rewrite rules are also strategies, i.e., functions satisfying the same type given above.
Let's look at a concrete type of programs, such as \Rise which is a pattern-based functional programming language where we want to apply well-known rewrite rules such as the fusion of two \inline{map} calls: $(map~f) \circ (map~g) \mapsto map~(f\circ g)$.
In \Rise, the left-hand side of the rule is expressed as:
\begin{lstlisting}[style=elevate-rise-short]
val p: Rise = fun(xs => map(f)(map(g)(xs)))
\end{lstlisting}
The AST representation of the body of this is shown in \cref{fig:mapFusion} on the left, with function applications explicit as \inline{app} nodes.

This is the implementation of the fusion rule in \Elevate:
\begin{lstlisting}[style=elevate-rise-short]
def mapFusion: Strategy[Rise] = p => p match {
  case app(app(map, f), app(app(map, g), xs)) =>
    Success( map(fun(x => f(g(x))))(xs) )
  case _ => Failure(mapFusion)                    }
\end{lstlisting}

Note that we are mixing expressions of the \Rise language (i.e., \inline{map(f)}) and \Elevate.
The expression nested inside \inline{Success} is the rewritten expression shown in \cref{fig:mapFusion} on the right.


\subsection{Strategy Combinators}
A key idea that \Elevate inherits from Stratego~\cite{visser2004program} is to describe strategies as compositions of other strategies.
Therefore, we introduce strategy combinators.

The \inline{seq} combinator is given two strategies \inline{fs} and \inline{ss} and applies the first strategy to the input program \inline{p}.
Afterwards, the second strategy is applied to the result.
\begin{lstlisting}[style=elevate-rise-short, basicstyle=\ttfamily\scriptsize]
def seq[P]: Strategy[P] => Strategy[P] => Strategy[P] =
  fs => ss => p => fs(p).flatMapSuccess(ss)
\end{lstlisting}
The \inline{seq} strategy is only successful when both strategies are successfully applied in succession, otherwise \inline{seq} fails.

In the implementation of \inline{seq}, we make use of the monadic interface of strategies:
the \inline{RewriteResult} ADT provides two versions of \lstinline[basicstyle=\ttfamily\footnotesize]{map/flatMap} to compose strategies -- one in case of a successful strategy application and one in case of failure.

The \inline{lChoice} combinator is given two strategies and applies the second strategy only if the first strategy failed.
\begin{lstlisting}[style=elevate-rise-short, basicstyle=\ttfamily\scriptsize]
def lChoice[P]: Strategy[P] => Strategy[P] => Strategy[P] =
  fs => ss => p => fs(p).flatMapFailure(_ => ss(p))
\end{lstlisting}

We use \inline{<+} as notation for \inline{lChoice} and \inline{`;`} for \inline{seq} and define two more combinators:

The \inline{try} combinator applies a strategy and in case of failure applies the identity strategy.
Therefore, \inline{try} never fails.
\begin{lstlisting}[style=elevate-rise-short, basicstyle=\ttfamily\scriptsize]
def try[P]: Strategy[P] => Strategy[P] =
  s => p => (s <+ id)(p)
\end{lstlisting}

\inline{repeat} applies a strategy until it is no longer applicable.
\begin{lstlisting}[style=elevate-rise-short, basicstyle=\ttfamily\scriptsize]
def repeat[P]: Strategy[P] => Strategy[P] =
  s => p => try(s `;` repeat(s) )(p)
\end{lstlisting}


\subsection{Traversal Strategies}
\begin{figure}
        \includegraphics[width=.65\linewidth]{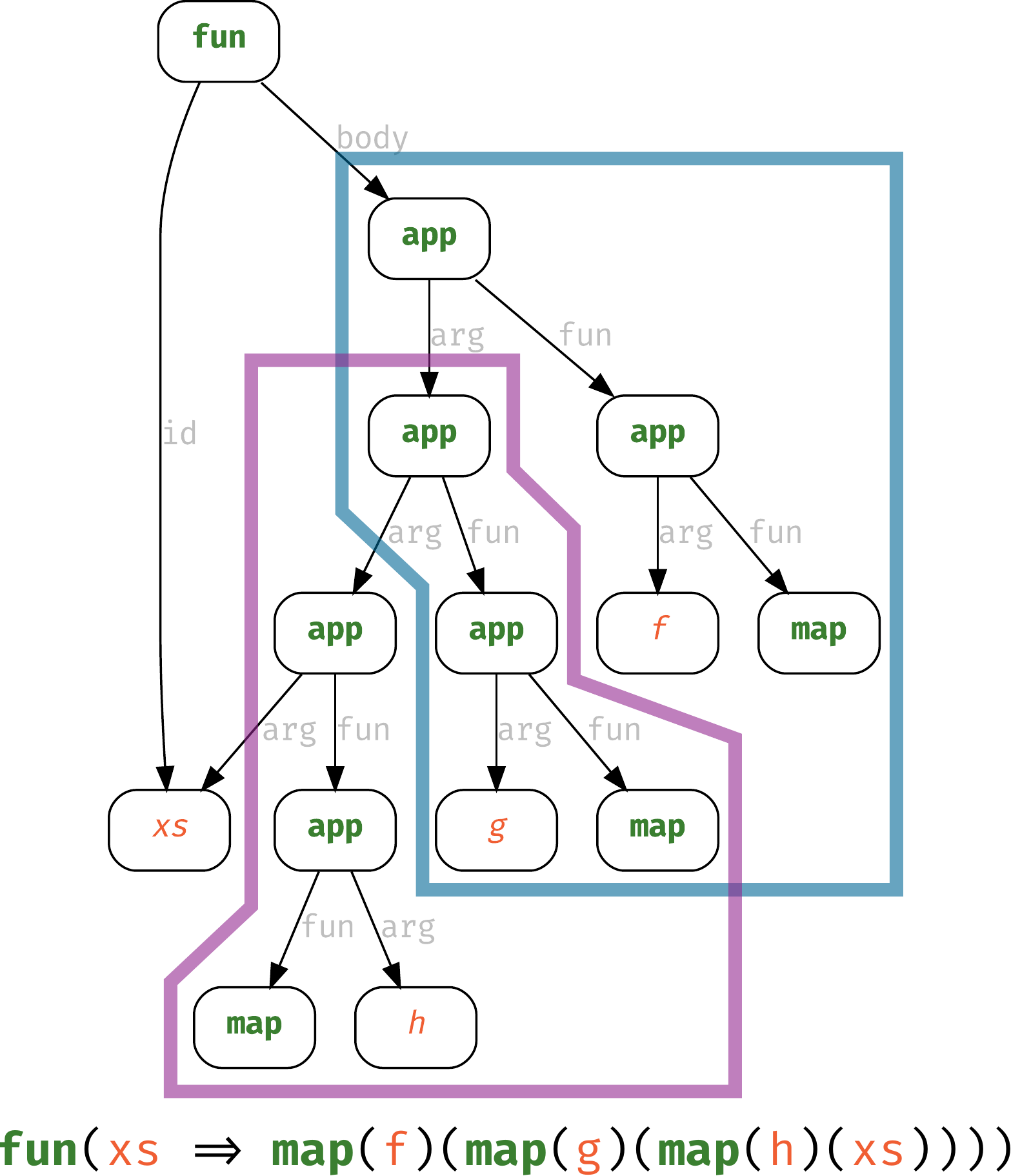}
        \caption{Two possible locations for applying the \emph{map-fusion} rule within the same program.}
        \label{fig:multipleLocations}
\end{figure}

The \inline{mapFusion} strategy we saw in the previous subsection is implemented as a function in \Elevate.
Therefore, its \inline{match} statement will try to pattern match its argument -- the entire program.
This means that a strategy on its own is very hard to reuse in different circumstances.

In addition, a strategy is often applicable at multiple places within the same program or only applicable at a specific location.
For example, the \inline{mapFusion} strategy is applicable twice in the following \Rise program:
\begin{lstlisting}[style=elevate-rise-short]
val threemaps = fun(xs => map(f)(map(g)(map(h)(xs))))
\end{lstlisting}
We may fuse the first or last two \inline{map}s as shown in \cref{fig:multipleLocations}.

In \Elevate, we use \emph{traversal strategies} to describe at which exact location a strategy is applied.
%
\citet{luttik1997specification} proposed three basic traversal strategies:
\begin{lstlisting}[style=elevate-rise-short]
def  all[P]: Strategy[P] => Strategy[P]
def  one[P]: Strategy[P] => Strategy[P]
def some[P]: Strategy[P] => Strategy[P]
\end{lstlisting}

\inline{all} applies a given strategy to all sub-expressions of the current expression and fails if the strategy is not applicable to all sub-expressions.
\inline{one} applies a given strategy to exactly one sub-expression and fails it the strategy is not applicable to any sub-expression.
\inline{some} applies a given strategy to at least one sub-expression but potentially more if possible.
\inline{one} and \inline{some} are allowed to non-deterministically choose sub-expressions.

In \Elevate, we see these three basic traversal strategies as a type class:
an interface that has to be implemented for each program type \inline{P}.
The implementation for \Rise is straightforward.
\Rise programs are represented by ASTs such as the one in \cref{fig:multipleLocations}, therefore, \inline{all}, \inline{one}, and \inline{some} correspond to the obvious implementations on the tree-based representation.


To fuse the first two \inline{map}s in \cref{fig:multipleLocations} we use the \inline{one} traversal strategy: \inline{one(mapFusion)(threemaps)}.
This will apply the \inline{mapFusion} strategy not at the root of the AST, but instead one level down first trying to apply the strategy (unsuccsessfully) to the function parameter and then (successfully) to the function body highlighted in the upper-right blue box.

To fuse the last two \inline{map}s we use the \inline{one} traversal strategy twice to apply \inline{mapFusion} two levels down in the AST: \inline{one(one(mapFusion))(threemaps)}.
This sucessfully applies the fusion strategy to the expression highlighed in the lower-left purple box in \cref{fig:multipleLocations}.


\subsection{Language-Specific Traversal Strategies}
The traversals we have discussed so far are not specific to a particular language, such as \Rise.
These traversals are flexible, but offer only limited control as for \inline{one} and \inline{some} the selection of sub-expressions is either non-deterministic, or implementation-dependent (as for \Rise) and in our context it makes rarely sense to apply a strategy to \inline{all} sub-expressions.


In \Elevate, one can easily specify program language specific traversal primitives.
\Rise is a functional language using $\lambda$-calculus as its representation.
Therefore, it makes sense to introduce traversals that navigate the two core concepts of $\lambda$-calculus: \lstinline[style=elevate-rise-short, basicstyle=\ttfamily\normalsize]{fun}ction abstraction and \lstinline[style=elevate-rise-short, basicstyle=\ttfamily\normalsize]{app}lication.

To apply a strategy to the body of a function abstraction we define the following traversal strategy:
\begin{lstlisting}[style=elevate-rise-short]
def body(s: Strategy[Rise]): Strategy[Rise] =
  p => p match                                       {
   case fun(x,b) => s(b).mapSuccess(nb => fun(x,nb))
   case _ => Failure(s)                              }
\end{lstlisting}

A strategy \inline{s} is applied to the function body and if successful a function is build around the transformed body.

Similarly we define traversals \inline{function} and \inline{argument} to traverse function applications:
\begin{lstlisting}[style=elevate-rise-short]
def function(s: Strategy[Rise]): Strategy[Rise] =
  p => p match                                       {
   case app(f,a) => s(f).mapSuccess(nf => app(nf, a))
   case _ => Failure(s)                              }

def argument(s: Strategy[Rise]): Strategy[Rise] =
  p => p match                                       {
   case app(f,a) => s(a).mapSuccess(na => app(f, na))
   case _ => Failure(s)                              }
\end{lstlisting}

For the \Rise program shown in \cref{fig:multipleLocations}, we are now able to precisely describe a traversal path in the AST.
To fuse the first two \inline{map}s we may write \inline{body(mapFusion)(threemaps)}, and to fuse the others \inline{body(argument(mapFusion))(threemaps)}.

The traversals defined here are specific to \Rise but similar traversals are obviously possible for any functional language.
If the program is not a functional language, say e.g., a computational graph as used by Tensorflow, different language-specific traversals (e.g., \inline{leftOperand} and \inline{rightOperand}) could be defined to describe language-specific traversals.




\subsection{Complete Expression Traversal Strategies}
All of the traversal primitives introduced so far apply their given strategies only to immediate sub-expressions.

Using strategy combinators and traversals, we are able to define recursive strategies which traverse entire expressions:
\begin{lstlisting}[style=elevate-rise-short]
def topDown[P](s: Strategy[P]): Strategy[P] =
  p => (s <+ one(topDown(s)))(p)
def bottomUp[P](s: Strategy[P]): Strategy[P] =
  p => (one(bottomUp(s)) <+ s)(p)
def allTopDown[P](s: Strategy[P]): Strategy[P] =
  p => (s `;` all(allTopDown(s)))(p)
def allBottomUp[P](s: Strategy[P]): Strategy[P] =
  p => (all(allBottomUp(s)) `;` s)(p)
def tryAll[P](s: Strategy[P]): Strategy[P] =
  p => (all(tryAll(try(s))) `;` try(s))(p)
\end{lstlisting}


\inline{topDown} and \inline{bottomUp} are useful strategies traversing an expression either from the top or from the bottom, trying to apply a given strategy at every sub-expression and stopping at the first successful application.
If the strategy is not applicable at any sub-expression, \inline{topDown} and \inline{bottomUp} fail.

\inline{allTopDown} and \inline{allBottomUp} do not use \inline{lChoice} insisting on applying the given strategy to every sub-expression.

The \inline{tryAll} strategy is often more useful as it wraps its given strategy in a \inline{try} and thus never fails but applies the strategy wherever possible.
Also note that the \inline{tryAll} strategy traverses the AST bottom-up instead of top-down.

These traversals have also been proposed by \citet{visser2004program} and we use them here with slightly different names more fitting for our use cases.

\subsection{Normalization, Confluence and Termination}
When implementing rewrite rules, such as the \inline{mapFusion} rule, as strategies, the match statement expects the program expression to be in a particular syntactic form.
For a functional language like \Rise, we might for example expect that expressions are fully $\beta$-reduced.
To ensure that expressions satisfy a \emph{normal form} we define:
\begin{lstlisting}[style=elevate-rise-short]
def normalize[P](s: Strategy[P]): Strategy[P] =
  p => repeat(topDown(s))(p)
\end{lstlisting}

The \inline{normalize} strategy applies a given strategy repeatedly at every possible sub-expression until it can not be applied any more.
Therefore, after \inline{normalize} successfully finishes it is not possible to apply the given strategy to any sub-expression any more.
By defining a strategy for $\beta$-reduction and using it together with \inline{normalize} we ensure that expressions are in $\beta$-normal-form.

Confluence (multiple non-deterministic rewrite paths eventually produce the same result) and termination are desirable properties for normal forms in term rewriting systems~\cite{visser2004program}.
In \Elevate, confluence only becomes a factor when the implementation of \inline{one} and \inline{some} are non-deterministic.
This can often be avoided such as for the use cases we consider with \Rise and \FSmooth that we will discuss in \cref{sec:case-study-1}.

Termination of normal forms critically depends on the chosen set of strategies.
Therefore, reasoning about terminating normal forms must be done on a case by case basis.
For example, it is trivial to build a non-terminating normal form using the \inline{id} strategy that is always applicable.
We currently, do not prevent the creation of non-terminating strategies similar as almost all general purpose computational languages do not prevent writing non-terminating programs.
In the future, we are interested to introduce a richer type system for \Elevate to better assist the user in writing well behaved strategies.

\subsection{Summary}
We have introduced \Elevate, a language for describing optimization strategies. 
In the next three sections we discuss three case studies of using \Elevate in the domains of automatic differentiation, image processing, and deep learning.

%% file: tex/fsmooth.tex
So far we have seen \Rise as the only example of a language that we transform with \Elevate, but \Elevate is flexible and not restricted to a single language.
In this first case study, we will look at the \FSmooth language that has been introduced in~\cite{DBLP:journals/pacmpl/ShaikhhaFVJ19}.

\FSmooth is a small functional language capable of automatically computing the derivative of arbitrary \FSmooth functions.
Implementing automatic differentiation is not too difficult but making it efficient is non-trivial.
\FSmooth achieves efficiency by rewriting the differentiated code.
In the paper, rewrite rules are specified alongside examples.
Example 5 in the paper shows that a \FSmooth program transposing a matrix twice can be rewritten into a program without transposition, see \cref{lst:fsmooth-example-5}.

\begin{figure}[b]
	\begin{subfigure}{\linewidth}
\begin{lstlisting}[style=fsmooth]
let MT = build (length M[0]) (fun i ->
          build (length M) (fun j -> M[j][i] ) ) in
         build (length MT[0])(fun i ->
          build (length MT) (fun j -> MT[j][i] ) )
\end{lstlisting}
	\end{subfigure}\vspace{-.75em}
	\begin{subfigure}{\linewidth}
\begin{lstlisting}[style=fsmooth]
build (length M) (fun i ->
 build (length M[0]) (fun j -> M[i][j] ) )
\end{lstlisting}
	\end{subfigure}
	\caption{Transposing a matrix twice in \FSmooth (top) and the rewritten program without transposition (bottom). From~\cite{DBLP:journals/pacmpl/ShaikhhaFVJ19}}
	\label{lst:fsmooth-example-5}
\end{figure}

The paper does not provide an explanation how the rewriting between these programs happens or is specified.
The authors only state \emph{``by applying the loop fusion rules and performing further partial evaluation the expression is derived''}.

We are interested in exploring the flexibility of \Elevate and if we can specify the rewrite rule applications programmatically.
We implemented the \FSmooth language representing expressions using an algebraic data type \inline{FSmooth}.
We implemented the rewrite rules from the paper, such as fusion rules:
\begin{lstlisting}[style=fsmooth-short]
(build e$_0$ e$_1$)[e$_2$] $\qquad\leadsto\quad$ e$_1$ e$_2$
length (build e$_0$ e$_1$) $\,\leadsto\quad$ e$_0$
\end{lstlisting}
These fusion rules are implemented as \Elevate strategies:
\begin{lstlisting}[style=elevate-rise-short, basicstyle=\ttfamily\scriptsize]
def buildGet(p: FSmooth): Strategy[FSmooth] = p match        {
 case app(get,(app(build,(e0,e1)),e2)) => Success(app(e1,e2))
 case _ =>                                Failure(buildGet)  }@\\[-.5em]@
def lenBuild(p: FSmooth): Strategy[FSmooth] = p match        {
 case app(length, app(build, (e0, e1))) => Success(e0)
 case _                                 => Failure(lenBuild) }
\end{lstlisting}

Using these rewrite rules encoded as \Elevate strategies, we use \inline{normalize} and \inline{lChoice} to specify multiple strategies that should be applied repeatedly at every sub-expression:
\begin{lstlisting}[style=elevate-rise-short]
normalize(buildGet <+ lenBuild <+ letPartialEval <+
          letApp <+ funToLet)(mt)
\end{lstlisting}

This \Elevate program successfully rewrites the doubly-transposed \FSmooth program into the non-transposed form.
By tracing the execution of the \Elevate program we get a sequence of 12 basic rewrite rule applications explaining the program transformation step-by-step.
A full version of the trace is shown in the supplementary material.


We also implemented the other examples in the paper which contain rewriting in \Elevate -- and identifying a minor bug in the description of example 6.

This case study shows \Elevate's flexibility to implement an existing rewrite system.
It also demonstrates \Elevate's ease of use:
we did not have to think about where to apply the individual rewrite rules thanks to the abstraction provided by \inline{normalize}.
It is important to stress, that this is not a built-in abstraction but defined itself in terms of the smaller building blocks \inline{repeat} and \inline{topDown}.
In the next case-study, we discuss how the ability to leverage these abstractions and build custom ones enables the definition of optimization strategies for image processing.

%% file: tex/halide.tex
In this case study, we are interested to see how our extensible strategy language \Elevate compares directly to Halide.
We will look at the binomial filter example which we already briefly showed in \cref{lst:halide-bf} and \ref{lst:elevate-bf}.
We will use \Rise as our computational language.
\Rise is a \Lift~\cite{DBLP:conf/icfp/SteuwerFLD15}-like language that is compiled by rewriting high-level pattern-based programs into a low-level representation encoding implementation and optimization choices.
It then uses a compilation process similar to~\cite{DBLP:journals/corr/abs-1710-08332} to compile the low-level pattern-based code into parallel imperative code.

\subsection{Halide Schedules for the Binomial Filter}
For Halide, we saw two algorithms describing the computation of the binominal filter in \cref{lst:halide-bf}.
The naive version on the top is a straightforward implementation as a two-dimensional stencil.
The separated version on the bottom specifies the computation as a composition of vertical and horizontal one-dimensional stencils.
In Halide, it is not possible to take the naive algorithm and use a schedule to specify the separability optimization.
When we do not specify a schedule in Halide, a default schedule is chosen that inlines computations as much as possible.
Therefore, the naive and the separated version with the default schedule result in code executing two nested loops and accessing 9 elements of the input image.
We can instruct Halide not to fully inline the vertical filter by writing: \halideInline{b_y.compute_at(out, y);}
This schedule uses the variable names from the Halide algorithm implementing the separated version of binomial filter.
The \halideInline{compute_at} schedule instruction tells the compiler to perform the computation of the vertical filter (\halideInline{b_y}) inside the for loop iterating over the \halideInline{y} dimension of the \halideInline{out} image.
This computation will be stored in a temporary that is consumed in the nested loop iterating over the \halideInline{x} dimension of \halideInline{out}.
We call this version \emph{scanline} as the temporary stores an entire line of the image.

For all versions, it is possible to parallelize the outermost loop with: \halideInline{out.parallel(y);}







\subsection{\Elevate Strategies Optimizing Binomial Filter}
To compare with Halide, we express the binomial filter with \Rise as shown at the top of \cref{lst:elevate-bf}.
This formulation is the naive way to describe a two-dimensional filter in \Rise.
The filter is expressed using two-dimensional variations of the \inline{pad}, \inline{slide}, and \inline{map} high-level patterns in the style of~\cite{DBLP:conf/cgo/HagedornSSGD18}.
\inline{pad} models the boundary handling, the sliding window pattern \inline{slide} describes a neighborhood of values each of which is then processed by the \inline{map} pattern.
The \inline{2D} versions of these pattens are just macros defined as compositions of the one-dimensional versions and a few additional basic patterns.
The \inline{dot} product computation is also defined as a composition:
\begin{lstlisting}[style=elevate-rise-short]
dot(x)(y) = reduce(add)(0)(map(mult)(zip(x)(y)))
\end{lstlisting}

To express the separability as an \Elevate strategy we introduce an image-processing specific rewrite rule that describes how the dot product of the weights and the neighborhood inside the binomial filter is separated:
\begin{lstlisting}[style=elevate-rise-short,basicstyle=\ttfamily\scriptsize]
def separateDot(w2d:Rise, wh:Rise, wv:Rise): Strategy[Rise] =
 p => p match                                                {
  case app(app(app(reduce, @\underline{add}@), @\underline{0}@), app(app(map, @\underline{mult}@),
    app(app(zip, app(join, w)), app(join, nbh)))) if w==w2d =>
            Success(nbh |> map(dot(wh)) |> dot(wv))
  case _ => Failure(sepDot)                                  }
\end{lstlisting}
The underlined values in the pattern matching indicate that these must match for the pattern matching to succeed.

The strategy takes three parameters that are all expressions in the computational language \Rise.
\inline{w2d} represents the two-dimensional weights and \inline{wh} and \inline{wv} are the separated horizontal and vertical weights.
\Elevate does not automatically attempt to separate the weights nor does it attempt to prove that the horizontal and vertical weights are a valid separation of the two-dimensional weights.
These considerations are left to the user.
We aim to empower users to extend \Elevate which such domain-specific strategies.

To rewrite the naive binominal filter into the separated one, we apply \inline{separateDot} with \inline{topDown} as shown earlier as an \Elevate strategy at the bottom of \cref{lst:elevate-bf} resulting in:
\begin{lstlisting}[style=elevate-rise
  % ,float
  % ,label={lst:rise-separated}
  % , caption={Separated binomial filter.
  %          Rewriten from the navie version and with the strategy shown in \cref{lst:elevate-bf}.}
           ]
img |> pad2D(1) |> slide2D(3)(1) |> map2D(fun(nbh =>
  nbh |> map(dot(weightsH)) |> map(dot(weightsV)) ))
\end{lstlisting}

We express the lowering of high-level \Rise expressions with \Elevate strategies that encode low-level implementation choices such as whether we would like to parallelize or not.
In Halide, the built-in \halideInline{compute_at} primitive is used for the scanline version.
In \Elevate a user-defined strategy encodes the same transformation.
We will investigate more complex optimization and low-level implementation strategies in more detail in the third case study in \cref{sec:case-study-3}.

\subsection{Performance Evaluation}
Even though this paper discusses \Elevate and not \Rise we want to evaluate whether a rewrite-based approach by combining them is capable of achieving competitive performance compared to the industry-strength Halide compiler that uses more traditional compiler techniques.
We have seen that \Elevate allows the definition of optimization strategies in an extensible way and that this allows to express optimizations as strategies that are not expressible as Halide schedules (such as separability).
We are now interested to see if these strategies encoding the same optimization decisions lead to competitive performance when compiled with \Rise and compared to Halide.

\begin{figure}[b]
  \newcommand{\baseLabel}{naive}
  \newcommand{\baseParLabel}{naive-par}
  \newcommand{\separatedLabel}{\\[.75em]separated}
  \newcommand{\separatedParLabel}{\\[.75em]separated\\[-.25em]-par}
  \newcommand{\scanlineLabel}{scanline}
  \newcommand{\scanlineParLabel}{scanline-par}
  \begin{tikzpicture}
    \begin{axis}[
      width=0.45\textwidth,
      height=0.23\textwidth,
      ytick={0, 30, 60, 90, 120},
      ylabel=Runtime in ms,
      ymin=0,
      ybar=0pt,
      bar width=7pt,
      symbolic x coords={
        \baseLabel,
        \separatedLabel,
        \scanlineLabel,
        ,
        \baseParLabel,
        \separatedParLabel,
        \scanlineParLabel,
      },
      xtick=data,
      enlarge y limits={0.15,upper},
      legend style={at={(0.8,0.9)},anchor=north,legend columns=1,font=\small},
      y label style={font=\small},
      y tick label style={font=\small},
      x tick label style={rotate=0, font=\bfseries\scriptsize, align=center},
      cycle list/Paired,
      every axis plot/.append style={fill,draw=black,no markers}
    ]
    \addplot[fill = Paired-G] 
      coordinates {
        (\baseLabel,          101.463)
        (\baseParLabel,       24.9694)
        (\separatedLabel,     90.9039)
        (\separatedParLabel,  21.9345)
        (\scanlineLabel,      87.1208)
        (\scanlineParLabel,   23.9888)
         };

    \addplot[fill = Paired-F] 
      coordinates {
        (\baseLabel,          113.4875)
        (\baseParLabel,       27.1324)
        (\separatedLabel,     103.9173)
        (\separatedParLabel,  25.872)
        (\scanlineLabel,      97.1703)
        (\scanlineParLabel,   23.7972)
         };

    \legend{Halide, Elevate+Rise}
    \end{axis}
    \end{tikzpicture}
      \caption{Performance evaluation of Halide and \Rise generated code for the binomial filter application. Optimization decisions for \Rise are implemented as \Elevate strategies.}
    \label{lst:halide-rise-performance}
    \end{figure}
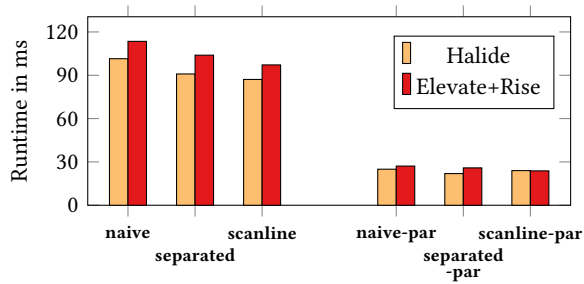

\Cref{lst:halide-rise-performance} shows the performance of the Halide and \Rise generated code measured on a ARM Cortex A7 quad-core\footnote{The 4 LITTLE cores of a Samsung Exynos 5 Octa 5422}.
We can see -- not surprisingly -- that the non-parallel versions on the left are significantly slower than the parallel versions.
The Halide generated code is 10-15\% faster than the \Rise generated code.
Improvements to the \Rise compiler might close this gap in the future.
Crucially, we observe the same trend for performance improvements due to optimizations.
This demonstrates that our extensible and rewrite based approach is capable of achieving competitive performance.

%% file: tex/tvm.tex
In our first two case studies, we looked at fairly simple strategies.
In our final case study, we are interested to explore the practicability and scalability of \Elevate.
We are looking at the domain of deep learning where performance optimizations are particularly important.
In this section, we explore the implementation of a scheduling language with \Elevate using the ability to define custom abstractions.
We use TVM~\cite{DBLP:conf/osdi/ChenMJZYSCWHCGK18} as an example for a state-of-the-art optimizing deep learning compiler with a scheduling API implemented in Python, similar to Halide.
We use \Rise again as the target language for the strategies we develop.

We start by looking at how TVM represents schedules and how we implement simple scheduling primitives such as \inline{parallel} and \inline{vectorize} in \Elevate.
Then we show how to implement more complex scheduling primitives like \inline{tile} using composition in \Elevate whereas it is built-in in TVM.
We follow a tutorial from the TVM authors\footnote{\url{https://docs.tvm.ai/tutorials/optimize/opt_gemm.html}} that discusses seven differently optimized versions of matrix multiplication.
For each one, we show an equivalent strategy implemented with \Elevate and evaluate the performance achieved.

\subsection{TVM Schedules for Matrix Multiply}
In TVM, computations are expressed similar to TensorFlow:
\begin{lstlisting}[style=tvm]
C = tvm.compute((M, N), lambda x, y:
  tvm.sum(A[x,k] * B[k,y], axis=k), name='C')
\end{lstlisting}

The \emph{``How to optimize GEMM on CPU''}\footnotemark[3] tutorial discusses seven versions applying different optimizations.
The \emph{baseline} version uses a default schedule with no instructions.
The \emph{blocking} version tiles the two outermost loops computing matrix $C$ and then splits the reduction loop before reordering the loop nest:
\begin{lstlisting}[style=tvm
  %,caption={TVM \emph{blocking} schedule}
  ]
# blocking version
xo, yo, xi, yi = s[C].tile(C.op.axis[0],C.op.axis[1],32,32)
k,             = s[C].op.reduce_axis
ko, ki         = s[C].split(k, factor=4)
s[C].reorder(xo, yo, ko, ki, xi, yi)
\end{lstlisting}

The \emph{vectorized} version vectorizes the innermost loop and, in addition, the \emph{loop permutation} version changes to a cache friendly access of matrix $A$ by reordering loops in a different way (switching \halideInline{ki} and \halideInline{xi}):
\begin{lstlisting}[style=tvm
  %,caption={TVM \emph{blocking} schedule}
  ]
# loop permutation version
xo, yo, xi, yi = s[C].tile(C.op.axis[0],C.op.axis[1],32,32)
k,             = s[C].op.reduce_axis
ko, ki         = s[C].split(k, factor=4)
s[C].reorder(xo, yo, ko, xi, ki, yi)
s[C].vectorize(yi)    # added already @\text{in}@ vectorized version
\end{lstlisting}

The \emph{array packing} version enables better memory accesses to matrix $B$ by introducing a temporary \halideInline{packedB} but requires changing the description of the computation in TVM:
\begin{lstlisting}[style=tvm
  %,caption={TVM \emph{blocking} schedule}
  ]
packedB = tvm.compute((N / 32, K, 32), lambda x, y, z:
  B[y, x * 32 + z], name='packedB')
C       = tvm.compute((M, N), lambda x, y:
  tvm.sum(A[x,k] * packedB[y // 32, k, tvm.indexmod(y,32)],
          axis=k), name='C')
\end{lstlisting}
With this rewritten computation we can now also describe the schedule affecting the computation of the temporary \halideInline{packedB} matrix.
Contrary to Halide, this computation is not inlined by default in TVM and results in a separate loop nest.
\begin{lstlisting}[style=tvm
  %,caption={TVM \emph{blocking} schedule}
  ]
# same as loop permutation version
x, y, z = s[packedB].op.axis
s[packedB].vectorize(z)
s[packedB].parallel(x)
\end{lstlisting}
The \emph{write cache for blocks} version allocates memory for the reduction accumulator and unrolls loops.
Building up on this, the fully \emph{parallel} version parallelizes the outermost loop:
\begin{lstlisting}[style=tvm
  %,caption={TVM \emph{blocking} schedule}
  ]
# parallel version
CC = s.cache_write(C, 'global')
xo, yo, xi, yi = s[C].tile(C.op.axis[0],C.op.axis[1],32,32)
xc, yc         = s[CC].op.axis
k,             = s[CC].op.reduce_axis
ko, ki         = s[CC].split(k, factor=4)
s[CC].reorder(ko, xc, ki, yc)
s[CC].unroll(ki)
s[CC].compute_at(s[C], yo)
s[CC].vectorize(yc)
s[C].parallel(xo)
x, y, z        = s[packedB].op.axis
s[packedB].vectorize(z)
s[packedB].parallel(x)
\end{lstlisting}

In the rest of the case study, we attempt to implement a similar looking scheduling language in \Elevate for rewriting \Rise programs.
We define abstractions ourselves similar to the built-in scheduling primitives provided by TVM.

\subsection{Basic Scheduling Primitives as \Elevate Strategies}
The TVM scheduling primitives \halideInline{parallel}, \halideInline{split}, \halideInline{vectorize}, and \halideInline{unroll} can be implemented as rewrite rules for \Rise.

\begin{lstlisting}[style=elevate-rise-short]
def parallel: Strategy[Rise] = p => p match          {
  case map => Success( mapPar )
  case _   => Failure( parallel )                    }
\end{lstlisting}
In \Rise, low-level implementation choices such as performing a computation in parallel are encoded with low-level patterns.
For example, the \inline{map} pattern that applies a function to each element of an array might be performed in a data-parallel fashion as indicated by the \inline{mapPar} variant of the pattern.
This is precisely what the \inline{parallel} strategy encodes: rewriting a \inline{map} pattern into its parallel variant.
A rewrite into the sequential variant \inline{mapSeq} is defined in the same style.

\begin{lstlisting}[style=elevate-rise-short]
def @\text{\color{RoyalPurple}{split}}@(n: Int): Strategy[Rise] = p => p match     {
  case app(map, f) =>
    Success( split(n) >> map(map(f)) >> join )
  case app(app(reduce, op), init) =>
    Success( split(n) >> reduce(fun(a => fun(y =>
      op(a, reduce(op)(init)(y)) )))(init) )
  case _ => Failure( @\text{\color{RoyalPurple}{split}}@(n) )                      }
\end{lstlisting}
TVM's \halideInline{split} scheduling primitive implements loop-blocking (also known as strip-mining).
In \Rise, this is achieved by transforming the computation over an array expressed by \inline{map(f)}:
first the input is split into an 2D array using \inline{split(n)}, then \inline{f} is \inline{map}ped twice to apply the function to all elements of the now nested array, and finally the resulting array is flattened into the original one-dimensional form using \inline{join}. 
We write \inline{f >>}~\inline{g} to indicate reverse function composition, i.e.,: \inline{fun(x =>}~\inline{g(f(x)))}.
It is important to note, that \Rise does not materialize the intermediate two-dimensional array in memory, but only uses this representation inside the compiler for code generation.
There is a second case for splitting \inline{reduce} resulting in two nested reductions.

\medskip
\begin{lstlisting}[style=elevate-rise-short]
def vectorize(n: Int): Strategy[Rise] = p => p match {
  case app(map, f) if isScalarFun(f) =>
    Success(asVector(n) >> map(mapVec(f)) >> asScalar)
  case _ => Failure( vectorize(n) )                  }
\end{lstlisting}
Vectorization is most efficient when applied to the innermost loop of a loop-nest.
In \Rise, this corresponds to applying the \inline{vectorize} strategy to the innermost \inline{map} of potentially nested \inline{map}s.
This is achieved in \Elevate by \inline{bottomUp(vectorize)}.
The extra constraint \inline{isScalarFun(f)} ensures that only functions operating on scalars are vectorized by inspecting \inline{f}'s type.
The restriction to scalar functions for vectorization is a current limitation of \Rise.

\medskip
\begin{lstlisting}[style=elevate-rise-short]
def unroll: Strategy[Rise] = p => p match            {
  case map    => Success( mapSeqUnroll )
  case reduce => Success( reduceSeqUnroll )
  case _      => Failure( unroll )                   }
\end{lstlisting}
The \inline{unroll} strategy rewrites the high-level \inline{map} and \inline{reduce} patterns into \Rise low-level patterns that will be unrolled by the \Rise compiler.

\paragraph{Identifying Locations}

In TVM, named identifiers describe the location at which the optimization should be applied.
For example, TVM's \inline{parallel} is invoked with an argument specifying the loop to parallelize.
Using named identifiers allows writing invalid schedules,\,e.g., trying to vectorize a reduction axis failing at runtime when TVM detects the error.

\Elevate does not use names to identify locations, but instead uses the traversals we defined in \cref{sec:elevate}.
The \inline{vectorize} strategy is -- by construction -- only applicable at valid locations within the AST.

By using \Elevate's traversal strategies, we can apply the basic scheduling strategies in a much more flexible way: e.g., \inline{topDown(parallel)} traverses the AST from top to bottom and will thus always parallelize the outermost \inline{map}, corresponding to the outermost for loop.
\inline{tryAll(parallel)} traverses the whole AST instead and all possible \inline{map}s are parallelized.
Depending on the desired use case users are free to combine different traversals with the basic scheduling strategies.

\subsection{Tiling as an \Elevate Strategy}
Tiling is an important optimization improving the cache hit rate by exploiting locality within a small neighborhood of elements.
TVM's \halideInline{tile} is a more complicated scheduling primitive to implement because it is essentially a combination of two traditional loop transformations: loop-blocking and loop-interchange.
In fact, \halideInline{tile} in TVM is a built-in combination of \halideInline{split} for loop-blocking and \halideInline{reorder} for loop-interchange.
We already saw how to implement \halideInline{split} using \Elevate and we will now discuss how to implement \inline{tile} using a combination of rules, normal-forms and domain-specific traversals.

We use matrix-matrix multiplication as the illustrative example to explain the optimization steps.
%
Since we use \Rise as the target language for the tiling strategy, we need to consider how matrix multiplication is expressed there:
\begin{lstlisting}[style=elevate-rise]
val dot = fun((a,b) => zip(a,b) |> map(*) |> reduce(+,0))
val mm = fun(a :: M.K.float => fun(b :: K.N.float =>
  map( fun(arow =>     // iterating over M
    map( fun(bcol =>     // iterating over N
      dot(arow, bcol)      // iterating over K
    )(transpose(b))
  )(a)
\end{lstlisting}
Essentially, the dot product is computed for each combination of rows and columns of matrix $A$ and $B$.


In the following, we show how to use \Elevate to systematically construct a strategy out of simple building blocks that has the same effect as TVM's \halideInline{tile} scheduling primitive.
Specifically, we construct a generalized strategy that is able to tile an arbitrary number of dimensions whereas TVM's \halideInline{tile} only tiles in two-dimensions.
\begin{lstlisting}[style=elevate-rise-short]
def tileND: List[Int] => Strategy[Rise]
\end{lstlisting}
This strategy expects a list of tile sizes, one per tiled dimension.
The two-dimensional tiling, that is equivalent to TVM's built-in \halideInline{tile} scheduling primitives, is expressed as \inline{tileND(List(x,y))(mm)} for this 2D case we also write \inline{tile(x,y)(mm)}.

The intuition for our \inline{tileND} implementation is simple:
First, we ensure that the required rules are applicable to the input expression by normalizing the expression using the \inline{DFNF} normal form.
Then, we apply the previously introduced \elevateUserFun{split} strategy to every \inline{map} to be blocked, recursively going from innermost to outermost.
Finally, we interchange dimensions accordingly.
\begin{lstlisting}[style=elevate-rise-short]
def tileND(n: List[Int]): Strategy[Rise] =
  DFNF `;` (n.size match {
    case 1 => function(split(n.head)) // loop-blocking
    case i =>
      fmap(tileND(d-1)(n.tail)) `;`   // recurse
      function(split(n.head))   `;`   // loop-blocking
      interchange(i)     })           // loop-reorder
\end{lstlisting}
In the following, we first introduce the required normal-forms (e.g., \inline{DFNF}), then explain how we recursively traverse (\inline{fmap}) to apply loop-blocking and finally briefly explain how we interchange dimensions in \Rise (\inline{interchange}).

\paragraph{Normal forms}
In \Elevate, we introduced \inline{normalize} to ensure that expressions are in a particular form expected by the implementation of rewrite rules.
$\lambda$-calculus (and \Rise) allows for semantically equivalent but syntactically different expressions.
For example, \inline{fun(x =>}~\inline{f(x))} is equivalent to \inline{f} \emph{iff} \inline{x} does not appear free in \inline{f}.
Transforming between these representations is called $\eta$-\emph{reduction} and $\eta$-\emph{abstraction}.

The simplest normal-form we use is the $\beta\eta$-\emph{normal-form} (\inline{BENF}) which exhaustively applies $\beta$- and $\eta$-reduction:
\begin{lstlisting}[style=elevate-rise-short]
def BENF = normalize(betaReduction <+ etaReduction)
\end{lstlisting}

As not every function abstraction is $\eta$-reducible, the function arguments of \Rise's higher-order patterns \inline{map} and \inline{reduce} might have different forms.
Therefore, we introduce a normal form making the data flow explicit by ensuring a function abstraction is present in every higher order pattern:
\begin{lstlisting}[style=elevate-rise-short]
def DFNF = BENF `;`
 // the argument of a map is a function abstraction
 normalize(argOf(map, not(isFun)`;`etaAbstraction))`;`
 // ... similar normalization for reduce primitive
\end{lstlisting}

The definition shows the normalization for \inline{map}.
A similar case exists for \inline{reduce}.
Using the \inline{not} and \inline{isFun} predicates, that are themselves \Elevate strategies, we describe the desired form in a natural and elegant way.

\paragraph{Recursively Applying Loop-Blocking}
In order to recursively apply the loop blocking strategy to nested \inline{map}s, we make use of the \Rise-specific traversal \inline{fmap}:
\begin{lstlisting}[style=elevate-rise-short, xleftmargin=.25\parindent]
def fmap(s:Strategy[Rise])=function(argOf(map,body(s)))
\end{lstlisting}
\inline{fmap} essentially traverses to the function argument of a \inline{map} primitive and applies the given strategy \inline{s}. For example,
\begin{lstlisting}[style=elevate-rise-short]
fmap(fmap(split(n)))(DFNF(map(map(map(f)))))
\end{lstlisting}
skips two \inline{map}s applying loop-blocking to the innermost \inline{map}.

\paragraph{Interchange in \texttt{tile}}
After recursively blocking all \inline{map}s, we use \inline{interchange} to rearrange the dimensions in the correct order.
For simplicity, we describe the two dimensional case:
after loop-blocking the data is four-dimensional and we must swap the two inner dimensions.
To achieve this we introduce two \inline{transpose} patterns and then move one of the \inline{transpose} into the right position.
Doing it this way every strategy is a \emph{semantics-preserving} transformation ensuring the correctness of the overall optimization.

\subsection{Reordering as an \Elevate Strategy}
Due to the loopless nature of \Rise, implementing TVM's \inline{reorder} primitive as a strategy is more complicated.
Instead of simply interchanging perfectly nested loops, the same is achieved in \Rise by interchanging the nesting of \inline{map} and \inline{reduce} patterns.
Therefore, there are multiple possible combinations to consider and the implementation of each rewrite rule requires reasoning about why exchanging the specific patterns is possible in the first place.


We implemented a \inline{reorder} strategy in \Elevate but its implementation is non-trivial and, therefore, it is not discussed here.
While it is possible to implement TVM's \halideInline{reorder} primitive, this particular loop transformation is just not a good fit for the pattern-based abstractions in the \Rise language.

\subsection{Matrix Multiply Schedules as \Elevate Strategies}
Using the TVM-like scheduling abstractions implemented as \Elevate strategies, we are now able to discuss how we combine them together to describe entire schedules in \Elevate.

For \emph{baseline}, TVM does not provide a schedule, but we describe the implicit behaviour of the compiler explicitly:
\begin{lstlisting}[style=elevate-rise]
(DFNF `;` topDown(fuseReduceMap) `;` lowerToC)(mm)
\end{lstlisting}
The TVM algorithm computes the dot product in a single statement.
The \Rise program describes the dot product with separate patterns which are fused using \inline{fuseReduceMap}.
The \inline{lowerToC} strategy lowers every high-level pattern into its low-level sequential version:
\inline{map} is rewritten into \inline{mapSeq} and \inline{reduce} into \inline{reduceSeq}.

For the \emph{blocking} version, we reuse the same \inline{lowerToC} strategy but first we leverage the abstractions that we have build emulating the TVM schedule in a similar style:
\begin{lstlisting}[style=elevate-rise]
val blocking = ( topDown(tile(32,32)) `;;`
                 topDown(isReduce `;` @\text{\color{RoyalPurple}{split}}@(4)) `;;`
                 topDown(reorder(Seq(1,2,5,6,3,4))) )
(blocking `;` lowerToC)(mm)
\end{lstlisting}
First we \inline{tile}, then we \elevateUserFun{split} and then we \inline{reorder}, just as specified in the TVM schedule.
We describe locations using the \inline{topDown} traversal and the \inline{isReduce} strategy predicate that applies the following \elevateUserFun{split} only if the current expression is a reduction.
We use the \inline{`;;'} combination to normalize using \inline{DFNF} between each step.


The \emph{loop permutation} version incorporates the changes of the \emph{vectorized} version by adding \inline{vectorize} and a different reordering of dimensions.
In contrast to TVM we identify dimensions by index rather than by name.
\begin{lstlisting}[style=elevate-rise]
val loopPerm = (topDown(tile(32,32)) `;;`
                topDown(isReduce `;` @\text{\color{RoyalPurple}{split}}@(4)) `;;`
                topDown(reorder(Seq(1,2,5,3,6,4)))) `;;`
                topDown(vectorize(32)))
(loopPerm `;` lowerToC)(mm)
\end{lstlisting}

For the \emph{array packing} version, we are not required to change the \Rise program manually, but can apply the array packing of matrix $B$ as a rewrite step.
Afterwards, we can reuse the \inline{loopPerm} strategy before the packed representation of $B$ is vectorized and then copied in parallel. 
\begin{lstlisting}[style=elevate-rise]
val arrayPacking = (packB `;;` loopPerm `;;`
                    topDown(vectorize(32)) `;;`
                    parallelizeCopy)
(arrayPacking `;` lowerToC)(mm)
\end{lstlisting}


For the \emph{parallel} version, we reuse the prior \emph{array packing} strategy only changing the way we lower the high-level code.
We parallelize the outermost loop with \inline{topDown(parallel)} and unroll the innermost reduction using \inline{bottomUp(isReduce `;` unroll)} before lowering the remaining high-level patterns to sequential code as before.
\begin{lstlisting}[style=elevate-rise]
(arrayPacking `;;` topDown(parallel) `;;`
 bottomUp(isReduce `;` unroll) `;` lowerToC)(mm)
\end{lstlisting}

We have demonstrated that it is feasible to implement a TVM-like scheduling language in \Elevate by expressing schedules as compositions of reusable strategies.

\subsection{Experimental Evaluation}
In order to evaluate the practicability and the scalability of \Elevate, we performed two experiments.

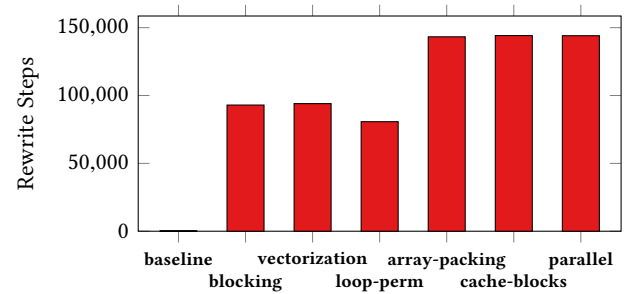
\begin{figure}[b]
  \newcommand{\baselineLabel}{baseline}
  \newcommand{\blockingLabel}{\\[.75em]blocking}
  \newcommand{\vectorizationLabel}{vectorization}
  \newcommand{\loopPermutationLabel}{\\[.75em]loop-perm}
  \newcommand{\arrayPackingLabel}{array-packing}
  \newcommand{\cacheBlocksLabel}{\\[.75em]cache-blocks}
  \newcommand{\parallelLabel}{parallel}
  \begin{tikzpicture}
    \begin{axis}[
      /pgf/number format/fixed,
      scaled y ticks = false,
      width=0.45\textwidth,
      height=0.25\textwidth,
      ylabel=Rewrite Steps,
      ymin=0,
      ybar=0pt,
      bar width=14pt,
      symbolic x coords={
        \baselineLabel,
        \blockingLabel,
        \vectorizationLabel,
        \loopPermutationLabel,
        \arrayPackingLabel,
        \cacheBlocksLabel,
        \parallelLabel,
      },
      xtick=data,
      y label style={font=\small},
      y tick label style={font=\small},
      x tick label style={rotate=0, font=\bfseries\scriptsize, align=center},
      cycle list/Paired,
      every axis plot/.append style={fill,draw=black,no markers}
    ]
    \addplot[fill = Paired-F]
      coordinates {
        (\baselineLabel,         211)
        (\blockingLabel,       92980)
        (\vectorizationLabel,  94030)
        (\loopPermutationLabel,80710)
        (\arrayPackingLabel,  143270)
        (\cacheBlocksLabel,   144183)
        (\parallelLabel,      144074)
    };
  
    \end{axis}
    \end{tikzpicture}
      \caption{Total number of successful rewrite steps when applying different optimization strategies.}
    \label{fig:rewrite-steps}
    \end{figure}

\paragraph{Number of Rewrite Steps}
In the first experiment, we are interested in the scalability of our approach by counting the number of successfully applied rewrites steps performed when applying a strategy to the \Rise matrix multiplication expression.
\Cref{fig:rewrite-steps} shows this number for every strategy shown in the previous subsection.
Since no major optimization strategies are applied to the \emph{baseline} version, only 211 rewrite steps are performed.
However, as soon as interesting optimizations are applied, we easily reach about 100,000 steps for the next three versions and about 150,000 for the most complicated optimizations.
The loop-permutation case slightly drops in numbers of applied rewrite rules because the specific nesting prescribed in TVM's schedule required fewer loop-interchanges.

These high numbers clearly show that abstractions are required to control this many rewrite steps.
It also shows the scalability of our compositional approach in which complex optimizations are composed out of a small set of fundamental building blocks.
The high-level strategies encode practical optimizations and hide massive numbers of individual rewrite steps that are actually performed.
Overall, applying the strategies to the \Rise expression took less than 50 seconds per version on a commodity notebook.

  \begin{figure}
  \newcommand{\baselineLabel}{baseline}
  \newcommand{\blockingLabel}{\\[.75em]blocking}
  \newcommand{\vectorizationLabel}{vectorization}
  \newcommand{\loopPermutationLabel}{\\[.75em]loop-perm}
  \newcommand{\arrayPackingLabel}{array-packing}
  \newcommand{\cacheBlocksLabel}{\\[.75em]cache-blocks}
  \newcommand{\parallelLabel}{parallel}
  \begin{tikzpicture}
    \begin{axis}[
      /pgf/number format/fixed,
      scaled y ticks = false,
      width=0.45\textwidth,
      height=0.23\textwidth,
      ylabel={Absolute Runtime (ms)},
      ymin=0,
      ybar=0pt,
      ymode=log,
      ytick={0, 50, 100, 200, 500, 1000, 2000},
      log ticks with fixed point,
      y tick label style={
          /pgf/number format/.cd,
              fixed,
              fixed zerofill,
              precision=1,
          /tikz/.cd
      },
      bar width=7pt,
      symbolic x coords={
        \baselineLabel,
        \blockingLabel,
        \vectorizationLabel,
        \loopPermutationLabel,
        \arrayPackingLabel,
        \cacheBlocksLabel,
        \parallelLabel,
      },
      xtick=data,
      legend style={at={(0.8,0.9)},anchor=north,legend columns=1,font=\small},
      y label style={font=\footnotesize},
      y tick label style={font=\small},
      x tick label style={rotate=0, font=\bfseries\scriptsize, align=center},
      cycle list/Paired,
      every axis plot/.append style={fill,draw=black,no markers}
    ]
    \addplot[fill = Paired-G] 
      coordinates {
        (\baselineLabel,         1961.470022)
        (\blockingLabel,       230.9482337)
        (\vectorizationLabel,  237.1613752)
        (\loopPermutationLabel,111.4528322)
        (\arrayPackingLabel,  109.1119637)
        (\cacheBlocksLabel,   99.76219681)
        (\parallelLabel,      71.86741923)
    };
  
   \addplot[fill = Paired-F] 
      coordinates {
        (\baselineLabel,      2742.235474)
        (\blockingLabel,       149.154503)
        (\vectorizationLabel,  149.1949995)
        (\loopPermutationLabel,111.581001)
        (\arrayPackingLabel,  108.184498)
        (\cacheBlocksLabel,   119.689003)
        (\parallelLabel,      41.785)
    };
  
    \legend{TVM, Elevate+Rise}
    \end{axis}
    \end{tikzpicture}
      \caption{Performance of TVM vs. \Rise generated code that has been optimized by \Elevate strategies.}
    \label{fig:tvm-elevate-performance}
    \end{figure}
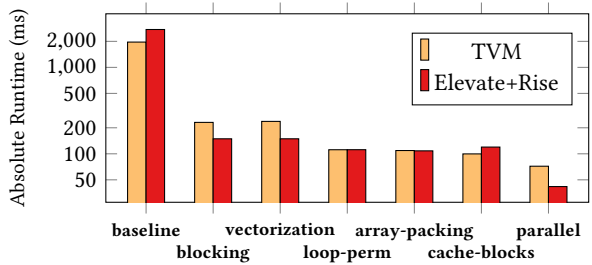

\paragraph{Performance Comparison}
In the second experiment, we are interested in the performance achieved when optimizing \Rise programs with \Elevate compared to TVM.
Ideally, the code optimized with \Elevate should be similar to the TVM-generated code and achieve competitive performance.
We generated LLVM code with TVM (version 0.6.dev) and C code for \Rise annotated with OpenMP pragmas for the version which include parallelization or vectorization.
The \Rise generated C code was compiled with clang (v.9.0.0) using \inline{-Ofast -ffast-math -fopenmp} which echoes the settings used by TVM and Halide\footnote{\url{https://github.com/halide/Halide/issues/2905}}.
The measurements were performed on an Intel core i5-4670K CPU (frequency locked to 3.4GHz) running Arch Linux (kernel 5.3.11-arch1-1).
We measured wall-time for \Rise-generated code and used TVM's built-in measurement API.
We measured 100 iterations per version reporting the median runtimes in milliseconds.

\Cref{fig:tvm-elevate-performance} shows the performance of \Rise and TVM generated code using a logarithmic scale.
The code generated by \Rise controlled by the \Elevate optimization strategies performs competitive to TVM.
Similar to the results of the Halide case study, our experiment shows a matching trend when comparing to TVM's versions with equivalent optimizations.
The most optimized parallel \Rise generated version improves the performance over the baseline by a factor of about 110x.
This means that the strategies we developed using \Elevate, that are defined in an extensible style by composing individual rewrite steps, scale to a level where they actually encode practically useable and relevant optimizations.

%% file: tex/related.tex
\paragraph{Term Rewriting and Strategy Languages}
\Elevate is inspired by existing strategy languages, especially ELAN~\cite{DBLP:journals/entcs/BorovanskyKKMV96,DBLP:journals/entcs/BorovanskyKKMR98} and Stratego~\cite{icfp/VisserBT98,visser2004program}, which introduce combinators to support user-defined strategies in the context of term rewriting systems.
Similar rewriting systems include~\cite{tcs/ClavelDELMMQ02, f-egc/PinaudAFKMV17,corr/abs-1102-2654,lopstr/FernandezKN11, cc/BrandDHJJKKMOSVVV01, DBLP:conf/ctrs/GoguenKKMMW87, boyle1997tampr}.
Program transformations using rewrite rules and strategy languages have since been used in many different domains including reverse engineering~\cite{DBLP:journals/software/ChikofskyC90}, refactoring~\cite{DBLP:books/daglib/0019908}, and obfuscation~\cite{DBLP:conf/popl/CollbergTL98}.
Visser~\cite{jsc/Visser05,tcs/Visser01} and Kirchner~\cite{birthday/Kirchner15} provide surveys covering term rewriting, strategy languages and their application domains.

To the best of our knowledge, \Elevate is the first strategy language that has been used to specify state-of-the-art compiler optimizations such as \inline{tiling} focusing on high performance code generation.

\paragraph{Rewriting in Compilers}
Rewrite rules and rewriting strategies have also been used when building compilers.
The Glasgow Haskell Compiler~\cite{peytonjones2001playing} uses rewrite rules as a practical way to optimize Haskell programs and Visser et. al.~\cite{icfp/VisserBT98} describe how to build program optimizers using rewriting strategies.
Other areas include building interpreters~\cite{entcs/DolstraV02}, instruction selection~\cite{rta/BravenboerV02} or constant propagation~\cite{entcs/OlmosV02}.
More recently, Lift~\cite{DBLP:conf/cgo/SteuwerRD17,DBLP:conf/cgo/HagedornSSGD18,DBLP:conf/icfp/SteuwerFLD15} showed how to use rewrite rules to generate high-performance code targeting accelerators.

Controlling the application of rewrite rules in compilers still largely is built-in in a fixed way based on heurisitcs.
In this work, we showed how to use \Elevate instead, allowing a more flexible and practical approach towards using rewrite rules for describing optimizations in compilers.

\paragraph{Schedule-based Compilers}
Halide~\cite{cacm/Ragan-KelleyASB18} introduced the concept of schedules describing program optimizations separate from the algorithm describing the computation.
This concept has been adopted by many other frameworks in domains including machine learning (TVM~\cite{DBLP:conf/osdi/ChenMJZYSCWHCGK18}), graph applications (GraphIt~\cite{DBLP:journals/pacmpl/ZhangYBKSA18}) or polyhedral compilation (Tiramisu~\cite{DBLP:conf/cgo/BaghdadiRRSAZSK19}, CHiLL~\cite{chen2008chill,hall2009loop}, AlphaZ~\cite{DBLP:conf/lcpc/YukiGKPR12}, URUK~\cite{DBLP:journals/ijpp/GirbalVBCPST06}).

These existing scheduling APIs are not designed as principled programming languages.
Instead, a fixed set of ad-hoc built-in primitives is exposed allowing users to specify which optimizations to apply.
In this work, we showed how to use \Elevate to implement scheduling languages from first principles as composition of rewrite rules.

%
%

%% file: tex/conclusion.tex
In this paper, we presented \Elevate{}: a language for describing optimization strategies.
\Elevate{} follows a tradition of prior systems used in different contexts that express optimization strategies as composition of rewrites.
We showed that, in contrast to existing systems with scheduling APIs such as Halide and TVM, programmers are not restricted to a set of built-in optimizations but define their own optimization strategies.
Using three case studies, we demonstrated \Elevate's \emph{flexiblity} to rewrite different languages (\FSmooth and \Rise), its \emph{extensibility} to add custom abstractions inlcuding domain-specific optimizations, and its \emph{practicality} to scale to complex optimization strategies requiring 150k rewrite steps used in deep learning.
We showed that \Elevate successfully optimizes programs in the \Rise language achieving competitive performance compared to Halide and TVM.